\documentclass[12pt]{iopart}
\usepackage[dvips]{graphicx}
\usepackage{verbatim}
\newcommand{\la}{\left\langle}
\newcommand{\ra}{\right\rangle}
\newcommand{\EPL}{Europhys.~Lett.~}
\newcommand{\MP}{Mol.~Phys.~}
\newcommand{\JCIS}{J.~Coll.~Int.~Sci.~}
\newcommand{\EPJ}{Eur.~Phys.~J.~}

\begin{document}

\title[Poisson-Boltzmann Theory: Limits of the Cell Model]
{Poisson-Boltzmann Theory of Charged Colloids:  Limits of the Cell Model
for Salty Suspensions}

\author{A. R. Denton\footnote[1]{Electronic address: {\tt alan.denton@ndsu.edu}}}
\address{Dept.~of Physics, North Dakota State University, Fargo, ND, U.S.A.
58108-6050}

\date{\today}

\begin{abstract}
Thermodynamic properties of charge-stabilised colloidal suspensions and polyelectrolyte
solutions are commonly modeled by implementing the mean-field Poisson-Boltzmann (PB) theory
within a cell model.  This approach models a bulk system by a single macroion, together with
counterions and salt ions, confined to a symmetrically shaped, electroneutral cell.
While easing numerical solution of the nonlinear PB equation, the cell model neglects
microion-induced interactions and correlations between macroions, precluding modeling of
macroion ordering phenomena.  An alternative approach, which avoids the artificial constraints
of cell geometry, exploits the mapping of a macroion-microion mixture onto a one-component
model of pseudo-macroions governed by effective interparticle interactions.
In practice, effective-interaction models are usually based on linear-screening approximations,
which can accurately describe strong nonlinear screening only by incorporating an effective
(renormalized) macroion charge.  Combining charge renormalization and linearized PB theories,
in both the cell model and an effective-interaction (cell-free) model, we compute
osmotic pressures of highly charged colloids and monovalent microions, in Donnan equilibrium
with a salt reservoir, over a range of concentrations.  By comparing predictions with
primitive model simulation data for salt-free suspensions, and with predictions of nonlinear
PB theory for salty suspensions, we chart the limits of both the cell model and linear-screening
approximations in modeling bulk thermodynamic properties.  Up to moderately strong electrostatic
couplings, the cell model proves accurate in predicting osmotic pressures of deionized
(counterion-dominated) suspensions.  With increasing salt concentration, however, the relative
contribution of macroion interactions to the osmotic pressure grows, leading predictions of
the cell and effective-interaction models to deviate.  No evidence is found for a
liquid-vapour phase instability driven by monovalent microions.  These results may guide
applications of PB theory to colloidal suspensions and other soft materials.
\end{abstract}


\maketitle

\section{Introduction}\label{intro}

When dispersed in water, or other polar solvents, colloidal particles can acquire
an electric charge through surface dissociation of counterions~\cite{evans99}.
Electrostatic repulsions between charged macroions, screened by surrounding microions
(counterions, salt ions), act to stabilise colloidal suspensions against aggregation
induced by attractive van der Waals forces, as first explained by the classic theory of
Derjaguin, Landau, Verwey, and Overbeek (DLVO)~\cite{DL,VO}.  Equilibrium and
dynamical properties of many soft and biologically prevalent materials, including
charge-stabilised colloids, polyelectrolytes, and ionic surfactants, are largely
determined by microion distributions and induced electrostatic interactions between
macroions.

Molecular simulations provide essentially exact results for model colloidal
suspensions and polyelectrolyte solutions, but can access only limited parameter ranges
of these complex systems.  For salt-free suspensions, recent Monte Carlo studies
of the primitive model have probed osmotic pressure, thermodynamic stability, and
macroion structure~\cite{linse99,linse-lobaskin99,lobaskin-linse99,linse00,
castaneda-priego06,brukhno09} and characterized effective macroion
interactions~\cite{stevens96,damico98,messina00,lobaskin-linse01,lobaskin-qamhieh03}
and bulk phase behaviour, including liquid-vapour separation~\cite{rescic-linse01,
hynninen05-critical,hynninen07} and melting~\cite{hynninen05-melting}.
At significant salt concentrations, proliferation of microions renders simulations
of the primitive model unwieldy, and data for bulk properties of salty suspensions
are scarce.

Materials with strong electrostatic interparticle interactions~\cite{interactions}
are commonly modeled via Poisson-Boltzmann (PB) theory~\cite{israelachvili92,deserno-holm01},
which combines the exact Poisson equation for the electrostatic potential with a
mean-field (Boltzmann) approximation for the microion densities.  Neglect of
correlations between microions is strictly justifiable only for weak electrostatic
couplings, characteristic of monovalent microions in water.  Nevertheless, the
mean-field PB theory often makes at least qualitatively correct predictions for
thermodynamic properties and provides a valuable reference for simulations and more
sophisticated theories that incorporate correlations.

Central to Poisson-Boltzmann theory is the PB equation governing the electrostatic potential.
For symmetric (isotropic) boundary conditions, this nonlinear differential equation is
easily solved --- analytically in planar and cylindrical geometries, and numerically
in spherical coordinates.  For this reason, PB theory is most often implemented within
a cell model~\cite{deserno-holm01,marcus55}, which reduces a bulk suspension to a single
macroion --- idealised as a charged sphere, rod, or plate --- centred in a cell of the
same shape, together with salt ions and neutralizing counterions.  While a symmetric cell
may reasonably approximate the Wigner-Seitz cell of a periodic crystal,
the computational advantage gained by imposing symmetry is offset by a loss of accuracy
in neglecting microion-induced interactions and correlations between macroions.
Although efficient for modeling nonlinear microion screening, the cell model is not
designed to describe macroion structure and ordering.

An alternative theoretical approach, which strives to incorporate macroion structure,
maps the multi-component ion mixture onto a one-component model by averaging over
microion degrees of freedom in the partition function~\cite{zvelindovsky07}.
The resulting pseudo-macroions are governed by an effective Hamiltonian involving
effective electrostatic interactions.  Although the one-component model is free of
the constraints of cell geometry, practical implementation usually requires
invoking linear-screening approximations, derived by linearizing the PB equation.
Moreover, in applying linearized theories to strongly-coupled suspensions of highly
charged macroions, it proves essential to consistently incorporate an effective
(renormalized) macroion charge to properly account for nonlinear microion screening~\cite{vongruenberg01,klein01,deserno02,tamashiro03}.

Condensation of counterions onto highly charged polyelectrolyte chains is well
established~\cite{manning69,oosawa71}.  By comparison, the association of counterions
with spherical macroions is a rather more subtle and less understood phenomenon.
Several theories have been proposed to describe counterion accumulation around highly
charged macroions and the resulting reduction of the effective macroion valence.
In a pioneering study, Alexander {\it et al.}~\cite{alexander84} defined an effective
macroion valence for the Debye-H\"uckel (linear-screening) theory by matching solutions
of the nonlinear and linearized PB equations at the boundary of a spherical cell.
In a series of studies, Levin and coworkers applied liquid-state theory~\cite{levin98,
tamashiro98,levin01} and derived an effective valence from a
jellium approximation~\cite{levin03,trizac-levin04,levin07,levin09} for the total free energy.
Other thermodynamic approaches are based on a thermal criterion for either the
electrostatic potential around a macroion~\cite{schmitz99,schmitz00,schmitz02}
or the effective pair potential~\cite{safran00}.  Recently, an effective valence
has been incorporated into the one-component model with linear-screening effective
interactions~\cite{zoetekouw_prl06,zoetekouw_pre06,denton08,lu-denton10}.

Effective macroion charges have been measured in experiments, e.g., by torsional
resonance spectroscopy~\cite{palberg92}, light or small angle neutron
scattering~\cite{gisler94,schurtenberger08}, and conductivity~\cite{palberg95}
measurements.  Renormalized charges also have been extracted from Monte Carlo (MC)
simulations according to various criteria:
equating the Debye-H\"uckel pressure~\cite{stevens96}
or chemical potential~\cite{tsao04} to the corresponding MC quantity;
applying the prescription of Alexander {\it et al.}~\cite{alexander84} to the
spherical cell model~\cite{lobaskin-linse99}; fitting an effective Yukawa pair potential
by inverting the macroion-macroion pair distribution function~\cite{lobaskin-linse01};
from the inflection point of the curve of accumulated running charge vs.
inverse separation from the macroion centre~\cite{lobaskin-qamhieh03,belloni98};
and from a dynamical criterion relating the counterion kinetic and potential
energies~\cite{diehl-levin04}.

The purpose of this paper is twofold: first, to generalize the charge renormalization
theory proposed in ref.~\cite{denton08} from salt-free suspensions to suspensions
dialyzed against an electrolyte reservoir; second, to apply the generalized theory
to salty suspensions to test the accuracy of the PB cell model in predicting osmotic pressure
over a wide range of salt concentrations.  By comparing results with available simulation data
for deionized suspensions and with predictions of nonlinear PB theory for salty suspensions,
we explore the limits of both the cell model and linearization approximations in predicting
bulk thermodynamic properties.  Up to moderately strong electrostatic couplings, the
cell model proves accurate in predicting osmotic pressures of counterion-dominated suspensions
(low salt concentrations).  With increasing salt concentration, as macroion interactions
gain in relative importance, predictions of the one-component (effective-interaction) model
increasingly deviate from those of the cell model.  No evidence is found of a liquid-vapour
spinodal instability for a physically rational definition of the renormalization threshold.

The paper is organized as follows.  Section~\ref{models} first reviews the relevant models:
the primitive model of charged colloids, the charge renormalization model for the effective
macroion valence, and the one-component model governed by an effective Hamiltonian.
Working from the density-functional formulation of Poisson-Boltzmann theory,
Sect.~\ref{pb} then derives the PB equation and outlines the two main implementations:
the cell model and the effective-interaction model.  In Sect.~\ref{linearization},
the two implementations of PB theory are linearized about the mean potential or, equivalently,
the mean microion densities.  Section~\ref{results} applies the nonlinear and linearized
PB theory to calculate the osmotic pressure and compares predictions with available data
from primitive model simulations of salt-free suspensions.
Finally, Sect.~\ref{conclusions} summarizes and concludes.

\section{Models}\label{models}

\subsection{Primitive Model}\label{pmsec}

The primitive model of charged colloids and polyelectrolytes idealises the solvent as a
homogeneous dielectric continuum of relative permittivity $\epsilon$.  Dispersed throughout
the solvent are macroions and microions, here modeled, respectively, as charged hard spheres
of radius $a$ and valence $Z$ (charge $-Ze$) and monovalent point ions.
In a closed suspension, all particles are confined to the same volume $V$.
In Donnan equilibrium, only the macroions are confined, while the microions
can exchange (e.g., across a semi-permeable membrane) with a salt reservoir, here assumed
to be a 1:1 electrolyte.  Modeling the reservoir as an ideal gas of ions of pair number
density $n_0$, the suspension has fixed salt chemical potential $\mu_s=2\mu_0$, where
$\mu_0=k_BT\ln(n_0\Lambda^3)$ is the chemical potential of each microion species at
absolute temperature $T$, the thermal wavelength $\Lambda$ defining the arbitrary zero
of the chemical potential.

\begin{comment}
The primitive model provides a simple, if somewhat limited, representation of real colloidal
suspensions and polyelectrolyte solutions.  The neglect of solvent molecular structure
clearly prohibits description of phenomena on microscopic ($< 1$ nm) length scales and
corresponding time scales.  Moreover, to avoid complexities of image charges and
polarization effects, index-matching of macroions to solvent must be assumed.
Finally, although a near monodisperse size distribution is experimentally achievable,
charge polydispersity is less controllable for colloidal macroions.  Despite its simplicity,
this useful reference model captures many aspects of soft materials and allows testing of
theories by comparing predictions with simulation data for the same model.
\end{comment}

A bulk suspension of $N_m$ macroions, $N_c$ counterions, and $N_s$ dissociated pairs
of oppositely charged salt ions contains $N_+=N_c+N_s$ positive and $N_-=N_s$
negative microions, for a total of $N_{\mu}=N_c+2N_s$ microions.
Global electroneutrality constrains the macroion and counterion numbers by the
condition $ZN_m=N_c$.  Accounting for excluded-volume and electrostatic (Coulomb)
pairwise interparticle interactions, the primitive model Hamiltonian can be expressed as
\begin{equation}
H=H_m+H_{\mu}+H_{m\mu}~,
\end{equation}
which separates naturally into three terms: a macroion Hamiltonian,
\begin{equation}
H_m=H_{\rm hs}+\frac{Z^2e^2}{2\epsilon}\sum_{{i\neq j=1}}^{N_m}\frac{1}{r_{ij}}~,
\label{Hm}
\end{equation}
where $H_{\rm hs}$ is the macroion hard-sphere Hamiltonian (including kinetic energy)
and $r_{ij}$ is the centre-centre separation between ions; a microion Hamiltonian,
\begin{equation}
H_{\mu}=K_{\mu}+\frac{e^2}{2\epsilon}\sum_{{i\neq j=1}}^{N_{\mu}}\frac{z_iz_j}{r_{ij}}~,
\label{Hmu}
\end{equation}
with microion valences $z_i=\pm 1$ and kinetic energy $K_{\mu}$; and a
macroion-microion interaction energy,
\begin{equation}
H_{m\mu}=\frac{Ze^2}{\epsilon}\sum_{i=1}^{N_m}\sum_{j=1}^{N_{\mu}}\frac{z_j}{r_{ij}}~.
\label{Hmmu}
\end{equation}

\subsection{Charge Renormalization Model for Effective Charge}\label{CRM}

The parameter $Z$ in the primitive model corresponds to the maximum (bare) charge
of the macroions, measureable by titration.  In practice, this simple correspondence
may be complicated by chemical reactions at the macroion surface, dependent on pH and
salinity, which can affect ion dissociation~\cite{gisler94,belloni91,vongruenberg99}.
Even in the absence of charge regulation, however, the {\it effective} macroion charge
--- probed by electrophoresis, rheology, and scattering experiments --- may lie
considerably below the bare charge.

The distinction between bare and effective charges arises from the strong association
between counterions and highly charged macroions.  Planar and rodlike macroions
generate electrostatic potentials of sufficiently long range to condense counterions
above a threshold surface charge density~\cite{manning69}.  In contrast, the potential
outside of a spherical macroion decays too rapidly to overcome entropy and
irreversibly bind counterions.  Nevertheless, counterions that venture sufficiently
close to the macroion surface to become thermally (transiently) bound can be considered
to renormalize the bare charge to a lower effective charge~\cite{alexander84}.
The remaining, weakly associated, counterions contribute to screening of the
effective charge and are amenable to linear-screening approximations.

Akin to Oosawa's two-phase theory of polyelectrolyte solutions~\cite{oosawa71}, free and
bound counterions part ways at a distance from the macroion at which the deviation of
the electrostatic energy of counterion-macroion attraction from the average potential
energy rivals the typical thermal energy $k_BT$ per counterion.  Defining $\psi(r)$ as
the reduced electrostatic potential energy (in $k_BT$ units) at distance $r$ from the
centre of a macroion, the thickness of the spherical association shell $\delta$
(Fig.~\ref{fig-model}) is defined by the condition
\begin{equation}
|\psi(a+\delta)-\overline\psi|=C~,
\label{delta1}
\end{equation}
where $\overline\psi$ is the spatially-averaged potential and $C$ is a constant of
${\cal O}(1)$ yet to be specified.  At distances $r>a+\delta$, the deviation is
presumed to be sufficiently small that a linear-screening approximation is valid.
Assuming coions to be expelled from the association shell, the
``dressed" macroion~\cite{kjellander97},
consisting of a bare macroion and its shell of bound counterions, carries an
effective valence $\tilde Z\le Z$.
Statistical fluctuations in $\tilde Z$ are neglected in this simple model.
Other workers have applied a thermal criterion similar to Eq.~(\ref{delta1}) to the
electrostatic potential~\cite{schmitz99,schmitz00,schmitz02} or the
effective pair potential~\cite{safran00}.

\subsection{One-Component Model and Effective Hamiltonian}\label{ocm}

Simulations of the primitive model~\cite{linse00,castaneda-priego06} have characterized
bulk properties of salt-free (counterion-dominated) suspensions with relatively low
charge asymmetry ($Z<100$).  Significant concentrations of salt pose, however,
computational challenges for large-scale simulations of bulk suspensions.
Consider, for example, that at 10~mM salt concentration a suspension of 100 macroions
of radius $a=10$ nm at 1\% volume fraction contains ${\cal O}(10^6)$ particles,
all interacting via long-range Coulomb forces, and that the particle number scales as $a^3$.
Therefore, salt-dominated suspensions usually are modeled by first mapping the
multi-component mixture onto a one-component model (OCM).

In Donnan equilibrium, a suspension is governed by a semigrand partition function,
\begin{equation}
{\cal Z}=\langle\langle\exp(-\beta H)\rangle_{\mu}\rangle_m~,
\label{Z1}
\end{equation}
where $\la~\ra_{\mu}$ denotes a grand canonical trace over microion coordinates,
$\la~\ra_m$ a canonical trace over macroion coordinates, and $\beta\equiv 1/(k_BT)$.
The one-component mapping expresses the partition function in the form
\begin{equation}
{\cal Z}=\la\exp(-\beta H_{\rm eff})\ra_m~,
\label{Z2}
\end{equation}
where the effective Hamiltonian for a one-component system of pseudo-macroions
\begin{equation}
H_{\rm eff}\equiv H_m-k_BT\ln\langle\exp[-\beta(H_{\mu}+H_{m\mu})]\rangle_{\mu}
\equiv\Omega_{\mu}
\label{Omegamu}
\end{equation}
can be interpreted equivalently as the grand potential $\Omega_{\mu}$ of the microions
in the ``external" potential of the macroions.
For a closed suspension, with fixed microion numbers, $\la~\ra_{\mu}$ becomes a canonical
trace over microion coordinates and $\Omega_{\mu}$ is replaced by the Helmholtz free energy
of the microions.

Practical applications of the OCM require approximating $\Omega_{\mu}$.  For this purpose,
Poisson-Boltzmann theory is a powerful approach.  Below we briefly review PB theory and
two common implementations: the cell model and the effective-interaction approach.

\section{Poisson-Boltzmann Theory}\label{pb}

\subsection{Density-Functional Formulation}\label{dft}

Poisson-Boltzmann theory is most elegantly formulated within the framework of classical
density-functional theory of nonuniform fluids~\cite{lowen92,lowen_jcp93}.  Corresponding to
the primitive model Hamiltonian [Eqs.~(\ref{Hm})-(\ref{Hmmu})], there exists a Helmholtz
free energy functional $F[n_m({\bf r}), n_{\pm}({\bf r})]$, which (for a given external
potential) is a unique functional of the macroion and microion number density profiles,
$n_m({\bf r})$ and $n_{\pm}({\bf r})$, varying with spatial position ${\bf r}$.
This free energy functional separates, according to $F=F_{\rm id}+F_{\rm ex}+F_{\rm ext}$,
into a (purely entropic) ideal-gas free energy functional of all ions,
\begin{equation}
F_{\rm id}=k_BT\int_V{\rm d}{\bf r}\,\sum_{i=m,\pm}n_i({\bf r})\{\ln[n_i({\bf r})\Lambda^3]-1\}~,
\label{Fid}
\end{equation}
an excess free energy functional, $F_{\rm ex}=F_{\rm hs}+F_{\rm el}$, due to hard-sphere (hs)
and electrostatic (el) interparticle interactions, and a contribution $F_{\rm ext}$ due to
an external potential.  Neglecting interparticle correlations (mean-field approximation),
the electrostatic part of the excess free energy functional may be approximated as
\begin{equation}
F_{\rm el}=\frac{1}{2}\int_V{\rm d}{\bf r}\,\rho({\bf r})\Phi({\bf r})~,
\label{Fel}
\end{equation}
where $\rho({\bf r})=e[n_+({\bf r})-n_-({\bf r})-n_f({\bf r})]$ is the total
charge density, including the negative charge fixed on the macroion surfaces
of number density $n_f({\bf r})$, and
\begin{equation}
\Phi({\bf r})=\int{\rm d}{\bf r}'\,\frac{\rho({\bf r}')}{\epsilon|{\bf r}-{\bf r}'|}
\label{Phi}
\end{equation}
is the total electrostatic potential at position ${\bf r}$ due to all ions.
The potential and the total ion density are related via the exact Poisson equation
\begin{equation}
\nabla^2\Phi({\bf r})=-\frac{4\pi}{\epsilon}\rho({\bf r})~,
\label{Poisson1}
\end{equation}
which may be expressed in the form
\begin{equation}
\nabla^2\psi({\bf r})=-4\pi\lambda_B[n_+({\bf r})-n_-({\bf r})]~; \quad
\nabla\psi|_{\rm surface}=Z\lambda_B/a^2~,
\label{Poisson2}
\end{equation}
where $\psi\equiv\beta e\Phi$ is the reduced electrostatic potential,
$\lambda_B=\beta e^2/\epsilon$ defines the Bjerrum length, and the macroion charges
are absorbed into a boundary condition at the macroion surfaces.
The microion densities implicitly vanish inside the macroion cores.

For a given external potential, the equilibrium densities of all ions minimize the
total grand potential functional of the system.
Alternatively, fixing the macroion coordinates and regarding their charges as the source
of an external potential, the equilibrium microion densities alone minimize the microion
grand potential functional
\begin{equation}
\Omega_{\mu}[n_{\pm}({\bf r})]=F_{\mu}[n_{\pm}({\bf r})]-
\mu_+\int_V{\rm d}{\bf r}\,n_+({\bf r})-\mu_-\int_V{\rm d}{\bf r}\,n_-({\bf r})~,
\label{Omega1}
\end{equation}
defined as a Legendre transform of the microion free energy functional
\begin{equation}
F_{\mu}[n_{\pm}({\bf r})]=
k_BT\int_V{\rm d}{\bf r}\,\sum_{i=\pm}n_i({\bf r})\{\ln[n_i({\bf r})\Lambda^3]-1\}+
\frac{1}{2}\int_V{\rm d}{\bf r}\,\rho({\bf r})\Phi({\bf r})~,
\label{Fmu1}
\end{equation}
the microion (electro)chemical potentials $\mu_{\pm}$ being identified as the Legendre
variables.  Note that $\Omega_{\mu}$ depends parametrically on the macroion coordinates
and that $F_{\mu}$ includes macroion-macroion Coulomb interactions for electroneutrality.
Under the assumption that either the electrostatic potential or the electric field vanishes
everywhere on the boundary of the volume $V$, the microion free energy functional also may be
expressed in the form
\begin{equation}
\beta F_{\mu}[n_{\pm}({\bf r})]=\int_V{\rm d}{\bf r}\,
\sum_{i=\pm}n_i({\bf r})\{\ln[n_i({\bf r})\Lambda^3]-1\}
+\frac{1}{8\pi\lambda_B}\int_V{\rm d}{\bf r}\,|\nabla\psi|^2~.
\label{FPB}
\end{equation}
Minimizing $\Omega_{\mu}[n_{\pm}({\bf r})]$ with respect to $n_{\pm}({\bf r})$
now yields the mean-field Boltzmann approximation for the equilibrium microion densities
\begin{equation}
n_{\pm}({\bf r})=n_{\pm}^{(0)}\exp[\mp\psi({\bf r})] \quad ({\rm fixed~macroions})~,
\label{boltzmann}
\end{equation}
the reference densities, $n_{\pm}^{(0)}=\Lambda^{-3}\exp(\beta\mu_{\pm})$, being the
microion densities at the reference potential $\psi=0$.  Evaluating the microion
grand potential functional [Eq.~(\ref{Omega1})] at the equilibrium microion density
profiles [Eq.~(\ref{boltzmann})] yields the microion grand potential
\begin{equation}
\beta\Omega_{\mu}=-\int_V{\rm d}{\bf r}\,[n_+({\bf r})+n_-({\bf r})]
-\frac{1}{2}\int_V{\rm d}{\bf r}\,[n_+({\bf r})-n_-({\bf r})
+n_f({\bf r})]\psi({\bf r})~.\quad
\label{Omegaeq}
\end{equation}

For a closed suspension (fixed particle numbers), the chemical potentials of the two
microion species differ because of asymmetric interactions with the macroions:
$\mu_+\neq\mu_-$.  Correspondingly, the reference densities also differ:
$n_+^{(0)}\neq~n_-^{(0)}$.
In Donnan equilibrium, however, exchange of microions with a salt reservoir shifts the
intrinsic microion chemical potentials,
$\mu_{\pm}^{\rm in}=[\delta F_{\mu}/\delta n_{\pm}({\bf r})]_{\rm eq}$,
by the Donnan potential $\psi_D$:
\begin{equation}
\beta\mu_{\pm}=\beta\mu_{\pm}^{\rm in}\pm\psi_D=\beta\mu_0=\ln(n_0\Lambda^3)~.
\label{mur}
\end{equation}
The total chemical potentials, and so too the reference densities, of the two microion
species are thus equalized.  The equilibrium microion density profiles are then given by
\begin{equation}
n_{\pm}({\bf r})=n_0\exp[\mp\psi({\bf r})]~.
\label{boltzmann-donnan}
\end{equation}
The Donnan potential is interpreted physically as the change in electrostatic potential
across the reservoir-suspension interface, and mathematically as a Lagrange multiplier
for the constraint of global electroneutrality.

Combining the Poisson equation for the potential [Eq.~(\ref{Poisson2})] with the
Boltzmann approximation for the microion densities [Eq.~(\ref{boltzmann})
or (\ref{boltzmann-donnan})], the Poisson-Boltzmann equation takes the form
\begin{equation}
\nabla^2\psi({\bf r})=\left\{ \begin{array}
{l@{\quad}l}
-4\pi\lambda_B\left(n_+^{(0)}\exp[-\psi({\bf r})]-n_-^{(0)}\exp[\psi({\bf r})]\right)
& ({\rm closed}) \\[1ex]
\kappa_0^2\sinh\psi({\bf r})~; \quad \nabla\psi|_{\rm surface}=Z\lambda_B/a^2~,
& ({\rm Donnan})
\end{array} \right.
\label{PB2}
\end{equation}
where $\kappa_0=\sqrt{8\pi\lambda_B n_0}$ is the screening constant in the reservoir.
Beyond the boundary condition at the macroion surfaces, the boundary-value problem is
fully specified only by imposing another condition at the outer boundary of the system,
which depends on the practical implementation of the theory.

\subsection{Cell-Model Implementation}\label{cm}

The anisotropic boundary conditions on the nonlinear PB equation imposed by an
arbitrary configuration of macroions render a general solution of Eq.~(\ref{PB2})
computationally daunting.  In recent years, powerful {\it ab initio} methods
have been developed for combining PB theory of microion density profiles with
molecular dynamics~\cite{lowen92,lowen_jcp93,lowen_el93,tehver99} or Brownian
dynamics~\cite{dobnikar03,dobnikar04} simulation to evolve macroion coordinates
according to derived forces.  Despite such advances, however, most applications
of PB theory have been implemented within a cell model to facilitate numerical
solution.

In a seemingly bold reduction, the cell model represents a bulk suspension by a
single macroion, neutralizing counterions, and salt ions confined to a cell
of the same shape as the macroion.  Microion-induced interactions between macroions
are simply ignored.  For spherical colloids, the natural choice is a spherical cell
centred on the macroion~\cite{note-eccentric}.  Gauss's law then dictates that the
electric field must vanish everywhere on the boundary of the electroneutral cell.
With the potential and microion densities depending on only the radial coordinate
$r$, the PB equation reduces to an ordinary differential equation for $\psi(r)$
with boundary conditions $\psi'(a)=Z\lambda_B/a^2$ and $\psi'(R)=0$,
where the cell radius $R$ is commensurate with the average macroion density:
$n_m=N_m/V=3/(4\pi R^3)$.  For a closed suspension, the arbitrary location
of the reference point of the electrostatic potential (where $\psi=0$) is usually
chosen as the cell boundary: $\psi(R)=0$.  In Donnan equilibrium, the potential
is conventionally chosen to vanish in the reservoir, in which case the boundary value
of the electrostatic potential is identified as the Donnan potential:
$\psi(R)=\psi_D\neq~0$.

An appealing feature of the cell model is the simple analytic relation between the
bulk pressure $p$ and the microion densities at the cell boundary:
\begin{equation}
\beta p=n_+(R)+n_-(R)~.
\label{pressure}
\end{equation}
Although first derived within the mean-field PB framework~\cite{marcus55},
this pressure relation proves to be exact within the
cell model~\cite{wennerstrom82}, i.e., valid also for correlated microions.
In Donnan equilibrium with an ideal-gas reservoir, the osmotic pressure $\Pi$,
i.e., the difference in pressure between suspension and reservoir,
is then given by
\begin{equation}
\beta\Pi=n_+(R)+n_-(R)-2n_0~.
\label{osmotic-pressure}
\end{equation}
Within PB theory, the osmotic pressure is strictly positive~\cite{deserno02},
which follows simply from Eqs.~(\ref{boltzmann-donnan}) and
(\ref{osmotic-pressure}) and the inequality, $\cosh x>1$.

The cell model implements PB theory by approximating the microion grand potential
[Eq.~(\ref{Omegamu})] in the one-component mapping of the primitive model
(Sec.~\ref{ocm}).  Reducing a bulk suspension to a single macroion in a symmetrically
shaped cell with isotropic boundary conditions greatly facilitates solution of
the nonlinear PB equation.  The microion grand potential implicitly depends on the
average macroion density through the density-dependent cell radius.  Moreover,
bare Coulomb interactions among all macroions are included in an average sense
via the electroneutrality constraint.
The cell-model implementation excellently approximates the single-macroion
(one-body) contribution to the free energy and pressure.  The trade-off for fully
incorporating nonlinear microion screening, however, is complete neglect of
effective interactions and correlations among macroions induced by microions.
Accurately accounting for multi-macroion contributions to thermodynamic properties 
requires an effective-interaction implementation of PB theory.

\subsection{Effective-Interaction Implementation}\label{effint}

An alternative implementation of PB theory, also based on the one-component mapping,
focuses on effective interactions derived from perturbative expansion of the microion
grand potential about a uniform reference system, namely, a plasma of microions
unperturbed by the macroions.  By incorporating macroion interactions, this approach
can model both thermodynamic and structural properties of colloidal suspensions.
Several liquid-state theoretical frameworks have been developed.
Density-functional theories~\cite{vRH97,graf98,vRDH99,vRE99} expand
the ideal-gas free energy functional in a Taylor series in powers of deviations of
the microion density profiles from their mean values $\bar n_{\pm}$:
\begin{equation}
\beta F_{\rm id}[n_{\pm}({\bf r})]=\sum_{i=\pm}\left(N_i[\ln(\bar n_i\Lambda^3)-1]
+\frac{1}{2\bar n_i}\int{\rm d}{\bf r}\,[n_i({\bf r})-\bar n_i]^2+\cdots\right)~.
\label{Fid-DFT}
\end{equation}
Response theories~\cite{silbert91,denton99,denton00,denton04,denton06,denton07} expand the
microion density profiles about a reference plasma in powers of the macroion potential,
$\phi_{m\pm}({\bf r})=\int{\rm d}{\bf r}'\,v_{m\pm}(|{\bf r}-{\bf r}'|)n_m({\bf r}')$:
\begin{equation}
n_i({\bf r})=\bar n_i+\sum_{j=\pm}\int{\rm d}{\bf r}'\,\chi_{ij}(|{\bf r}-{\bf r}'|)
\phi_{mj}({\bf r}')+\cdots~,
\label{n-LRT}
\end{equation}
where $\chi_{ij}=\delta n_i({\bf r})/\delta\phi_{mj}({\bf r}')$ are linear response functions
of the reference plasma and higher-order terms involve nonlinear response functions.
Distribution function theories are based on density-functional expansions of the microion
correlation functions about a reference plasma~\cite{warren00,warren03,warren06,chan85,
chan-pre01,chan-langmuir01} or on integral-equation closures~\cite{patey80,belloni86,khan87,
carbajal-tinoco02,petris02,anta,outhwaite02}.

In all of these statistical mechanical schemes, insertion of the perturbative expansion
into Eq.~(\ref{Omegamu}) recasts the effective Hamiltonian (microion grand potential)
in the form of a sum of effective interactions:
\begin{equation}
H_{\rm eff}=E_v+H_{\rm hs}+\frac{1}{2}\sum^{N_m}_{i\neq j=1}v_{\rm eff}(r_{ij})+\cdots~,
\label{Heff2}
\end{equation}
where $E_v$ is a one-body volume energy -- independent of the macroion coordinates but
dependent on the average macroion density -- $v_{\rm eff}(r)$ is an effective electrostatic
pair potential summed over macroion pairs with centre-centre separation $r$, and
higher-order terms involve sums over effective many-body interactions.  The volume energy
here corresponds to the microion grand potential for a single macroion.

In practice, mean-field (Debye-H\"uckel-like) approximations for the electrostatic free energy,
response functions, or correlation functions, make the various effective-interaction approaches
essentially equivalent.  Thermodynamic and structural properties of a bulk system can be
calculated by inputting the effective interactions into statistical mechanical theories
or simulations of the OCM.

\section{Linearized Poisson-Boltzmann Theory}\label{linearization}

\subsection{Perturbation Expansion and Truncation}\label{perturbation}

Practical applications of PB theory often invoke a perturbation approximation,
whereby the microion densities on the right side of the PB equation [Eq.~(\ref{PB2})]
are expanded in powers of the deviation of the potential from a reference potential.
Although not unique, the reference potential is commonly chosen as the reservoir potential
($\psi=0$) for a suspension in Donnan equilibrium.  A more appropriate choice, however,
is the mean potential in the suspension $\overline\psi$, this being also the only consistent
choice for a closed suspension~\cite{vongruenberg01,klein01,deserno02,tamashiro03,denton07}.
The distinction between these two choices is especially relevant for deionized suspensions
of highly charged colloids.

A linear-screening approximation truncates the expansions of the microion densities
at first (linear) order in the potential.  Correspondingly, the microion grand potential
functional is expanded to quadratic order in the deviations of the microion densities
from their means~\cite{deserno02,tamashiro03}:
\begin{eqnarray}
\beta \Omega^{\rm lin}_{\mu}[n_{\pm}({\bf r})]&=&\sum_{i=\pm}\left\{N_i\left[\ln\left(
\frac{\bar n_i}{n_0}\right)-1\right]
+\frac{1}{2\bar n_i}\int{\rm d}{\bf r}\,[n_i({\bf r})-\bar n_i]^2\right\}
\nonumber\\
&+&\frac{1}{2}\int{\rm d}{\bf r}\,[n_+({\bf r})-n_-({\bf r})-n_f({\bf r})]\psi({\bf r})~,
\label{Omegamu-quadratic}
\end{eqnarray}
where $\bar n_{\pm}=N_{\pm}/V'$ here represent the average microion densities
in the free volume $V'$ (unoccupied by the macroion hard cores).
Minimizing the approximate functional with respect to $n_{\pm}({\bf r})$ now yields
the linearized microion density profiles [{\it cf}. Eq.~(\ref{boltzmann-donnan})]:
\begin{equation}
\ln\left(\frac{\bar n_{\pm}}{n_0}\right)+\frac{n_{\pm}({\bf r})-\bar n_{\pm}}{\bar n_{\pm}}
\pm\psi({\bf r})=0
\label{EL}
\end{equation}
and thus
\begin{equation}
n_{\pm}({\bf r})=\bar n_{\pm}[1\mp\psi({\bf r})\pm\overline\psi]~,
\label{npm-lin}
\end{equation}
where $\bar n_{\pm}=n_0\exp(\mp\overline\psi)$ for consistency.
Substituting Eq.~(\ref{npm-lin}) back into Eq.~(\ref{Omegamu-quadratic}) yields the
linearized microion grand potential per macroion:
\begin{equation}
\beta\omega^{\rm lin}_{\mu}=\sum_{i=\pm}x_i\left[\ln\left(
\frac{\bar n_i}{n_0}\right)-1\right]-\frac{Z}{2}\psi(a)+\frac{Z}{2}\overline\psi~,
\label{omegamu-lin-eq}
\end{equation}
where $x_{\pm}\equiv N_{\pm}/N_m$.
The first term on the right side originates from the microion entropy; the second term
combines the macroion self energy, $Z^2\lambda_B/(2a)$, and the macroion-counterion
interaction energy; and the final term accounts for the average potential energy of the microions
relative to the reservoir.  Practical implementations aim to calculate the potential.

\subsection{Cell-Model Implementation of Linearized Theory}\label{lcm}

Substituting the linearized microion densities [Eq.~(\ref{npm-lin})] into
[Eq.~(\ref{PB2})], the linearized PB equation in spherical coordinates takes the form
\begin{equation}
\psi''(r)+\frac{2}{r}\psi'(r)=\kappa^2[\psi(r)-\overline\psi]-4\pi\lambda_B(\bar n_+-\bar n_-)~,
\quad a\leq r\leq R~,
\label{LPB2}
\end{equation}
where $\kappa=\sqrt{4\pi\lambda_B(\bar n_++\bar n_-)}$ is the screening constant in the
suspension.  Note that $\kappa$ depends implicitly on the average macroion density via
the global constraint of electroneutrality and that $\kappa^2=\kappa_0^2\cosh\overline\psi$
in Donnan equilibrium, but not in closed equilibrium.
With cell boundary conditions, $\psi'(a)=Z\lambda_B/a^2$ and $\psi'(R)=0$, the
solution of Eq.~(\ref{LPB2}) is
\begin{equation}
\psi(r)=A_1\frac{e^{\kappa r}}{r}+A_2\frac{e^{-\kappa r}}{r}+A_3~, \quad a\leq r\leq R~,
\label{LPB-cell}
\end{equation}
where the coefficients are given by
\begin{equation}
A_1=Z\lambda_B\left[(\kappa a-1)e^{\kappa a}-(\kappa a+1)\frac{\kappa R-1}{\kappa R+1}
e^{\kappa(2R-a)}\right]^{-1}~,
\label{A1}
\end{equation}
\begin{equation}
A_2=A_1\frac{\kappa R-1}{\kappa R+1}e^{2\kappa R}~,
\label{A2}
\end{equation}
\begin{equation}
A_3=\overline\psi+\frac{\bar n_+-\bar n_-}{\bar n_++\bar n_-}~.
\label{A3}
\end{equation}
For a closed suspension, the choice $\psi(R)=0$ gives a mean potential
\begin{equation}
\overline\psi=-A_1\frac{2\kappa}{\kappa R+1}~e^{\kappa R}-\frac{\bar n_+-\bar n_-}
{\bar n_++\bar n_-}~.
\label{psibar-closed}
\end{equation}
For a Donnan suspension (with zero reservoir potential), averaging Eq.~(\ref{LPB-cell})
over the free volume of the spherical cell yields rather
\begin{equation}
\overline\psi\equiv\frac{3}{R^3-a^3}\int_a^R{\rm d}r\,r^2\psi(r)
=-\sinh^{-1}\left(\frac{3Z}{8\pi n_0(R^3-a^3)}\right)~.
\label{psibar-donnan}
\end{equation}

The pressure relation [Eq.~(\ref{pressure})], although exact in the cell-model
implementation of nonlinear PB theory, does not survive linearization.
In linearized PB theory, therefore, the pressure must be computed as a
thermodynamic derivative of the microion grand potential
[Eqs.~(\ref{Omegamu-quadratic}) and (\ref{omegamu-lin-eq})]:
\begin{equation}
p=-\left(\frac{\partial\Omega^{\rm lin}_{\mu}}{\partial V}\right)_{T,N_m,n_0}
=-\frac{1}{4\pi R^2}\left(\frac{\partial\omega^{\rm lin}_{\mu}}{\partial R}\right)_{T,n_0}~.
\label{pressure-omega}
\end{equation}

\subsection{Effective-Interaction Implementation of Linearized Theory}\label{leffint}

In contrast to the cell model, effective-interaction implementations of PB theory include
microion-induced interactions between macroions in the suspension.  Truncating the
ideal-gas free energy functional expansion [Eq.~(\ref{Fid-DFT})] at quadratic order, or
the microion density profile expansions [Eq.~(\ref{n-LRT})] at linear order, amounts to
neglecting three-body and all higher-order effective interactions.  At the mean-field level,
such linear-screening approximations predict an effective electrostatic pair potential
of Yukawa form,
\begin{equation}
\beta v_{\rm eff}(r)=Z^2\lambda_B\left(\frac{e^{\kappa a}}
{1+\kappa a}\right)^2\frac{e^{-\kappa r}}{r}~,\quad r\ge 2a~,
\label{veff}
\end{equation}
and a one-body volume energy [$E_v$ in Eq.~(\ref{Heff2})] identical in form to the
linearized microion grand potential [Eq.~(\ref{omegamu-lin-eq})].
The electrostatic potential differs, however, from the cell model prediction.
In contrast to the solution of the linearized PB equation in the spherical cell model
[Eqs.~(\ref{LPB-cell})-(\ref{A3})],
linear-response theory~\cite{denton99,denton00,denton04,denton06} and related
effective-interaction theories~\cite{vRH97,graf98,vRDH99} predict
\begin{equation}
\psi(r)=-Z\lambda_B~\frac{e^{\kappa a}}{1+\kappa a}
~\frac{e^{-\kappa r}}{r}~, \quad r\ge a~,
\label{psi0}
\end{equation}
with mean value
\begin{equation}
\overline\psi\equiv 4\pi\frac{N_m}{V'}\int_a^{\infty}{\rm d}r\, r^2\psi(r)
=-\frac{\bar n_+-\bar n_-}{\bar n_++\bar n_-}
=-\frac{3}{(\kappa a)^2}\frac{Z\lambda_B}{a}\frac{\eta}{1-\eta}
\label{psibar-open}
\end{equation}
for macroion volume fraction $\eta=(4\pi/3)a^3(N_m/V)=1-V'/V$.  This potential is
identical to the solution of the linearized PB equation for a single macroion in a bulk
suspension under open boundary conditions: $\psi(r)$ and $\psi'(r)\to 0$ as $r\to\infty$.

Taking care to consistently incorporate density dependence, the effective electrostatic
interactions (including the volume energy) may be input into simulations~\cite{lu-denton07}
or statistical mechanical theories for the total semigrand potential
\begin{equation}
\Omega=-k_BT\ln{\cal Z}=E_v+F_m~.
\label{Omega}
\end{equation}
In linear-response theory, the macroion free energy
\begin{equation}
F_m=-k_BT\ln\la\exp\left(-\beta H_{\rm hs}-\frac{1}{2}\sum^{N_m}_{i\neq j=1}
\beta v_{\rm eff}(r_{ij})\right)\ra_m~,
\label{Fm1}
\end{equation}
which includes contributions from the macroion entropy and interactions,
can be reasonably approximated by a simple variational approach.  First-order
thermodynamic perturbation theory with a hard-sphere reference system~\cite{HM}
predicts a macroion free energy per macroion
\begin{equation}
f_m=\min_{(d)}\left\{f_{\rm hs}(d)+2\pi n_m\int_d^{\infty}{\rm d}r\,
r^2g_{\rm hs}(r;d)v_{\rm eff}(r)\right\}~,
\label{Fm2}
\end{equation}
where $f_{\rm hs}$ and $g_{\rm hs}$ are the free energy per macroion and radial
distribution function, respectively, of the reference system.  Minimization with
respect to the effective hard-sphere diameter $d$ generates a least upper bound
to $f_m$ .

\subsection{Charge Renormalization Theory}\label{CRT}

Linearized PB theory is valid only for bare macroion charges sufficiently low that
nonlinear microion screening can be safely neglected.
One might conceive of modeling a highly charged suspension simply by combining the
cell model solution of the nonlinear PB equation
with the linear-screening approximation for the macroion effective pair potential
[Eq.~(\ref{veff})].
However, without consistently incorporating nonlinear corrections to the pair
and many-body effective interactions~\cite{denton04,denton06}, such a hybrid
approximation would be uncontrolled.

A more consistent and computationally practical synthesis of nonlinear screening and
effective interactions invokes the charge renormalization model of Sec.~\ref{CRM}
to define an effective macroion valence $\tilde Z$ within the linear regime.  In this scheme,
only the $\tilde N_{\pm}$ free microions (with $\tilde N_+-\tilde N_-=\tilde ZN_m$),
are described by a linear-screening approximation, the remaining $Z-\tilde Z$ bound counterions
(per macroion) renormalizing the bare charge.
A theory developed previously for salt-free suspensions~\cite{denton08,lu-denton10}
is here generalized to salty suspensions.

The electrostatic potential around a dressed macroion of effective valence $\tilde Z$ and
association shell thickness $\delta$ is given from Eq.~(\ref{psi0}) by
\begin{equation}
\tilde\psi(r)=-\tilde Z\lambda_B~\frac{e^{\tilde\kappa(a+\delta)}}{1+\tilde\kappa(a+\delta)}
~\frac{e^{-\tilde\kappa r}}{r}, \quad r\ge a+\delta~,
\label{psir}
\end{equation}
with mean value [{\it cf.} Eq.~(\ref{psibar-open})]
\begin{equation}
\la\tilde\psi\ra=-\frac{\tilde n_+-\tilde n_-}{\tilde n_++\tilde n_-}~.
\label{psibar-open-cr}
\end{equation}
Here $\tilde n_{\pm}=\tilde N_{\pm}/[V(1-\tilde\eta)]$ are mean number densities of
free microions, $\tilde\eta=\eta(1+\delta/a)^3$ is the effective volume fraction of the
dressed macroions, and $\tilde\kappa=\sqrt{4\pi\lambda_B(\tilde n_++\tilde n_-)}$
is a renormalized screening constant.
From Eqs.~(\ref{delta1}) and (\ref{psir}), the association shell thickness is now specified
by solving
\begin{equation}
\left|\frac{\tilde Z\lambda_B}{[1+\tilde\kappa(a+\delta)](a+\delta)}+\la\tilde\psi\ra\right|=C
\label{delta2}
\end{equation}
self-consistently for $\delta$ as a function of $\tilde Z$, noting that $\tilde\kappa$
depends implicitly on $\tilde Z$ and $\delta$.

The basic distinction between free and bound counterions implies a corresponding
separation of the volume energy per macroion:
\begin{equation}
\varepsilon_v=\omega_{\rm free}+f_{\rm bound}~.
\label{epsilonv}
\end{equation}
The linear-screening approximation is applied only to the free microions, whose
grand potential (per macroion) is approximated by
\begin{equation}
\beta\omega_{\rm free}=
\sum_{i=\pm}\tilde x_i\left[\ln\left(\frac{\tilde n_i}{n_0}\right)-1\right]
-\frac{\tilde Z^2}{2}~\frac{\tilde\kappa\lambda_B}{1+\tilde\kappa(a+\delta)}
-\frac{\tilde Z}{2}~\frac{\tilde n_+-\tilde n_-}{\tilde n_++\tilde n_-}~,
\label{omega-free}
\end{equation}
where $\tilde x_{\pm}\equiv\tilde N_{\pm}/N_m$.  Note that $\omega_{\rm free}$
has the same form as $\omega_{\mu}^{\rm lin}$ [Eq.~(\ref{omegamu-lin-eq})], except that
bare parameters are replaced by renormalized parameters and the macroion self energy
is omitted, being now associated with the bound counterions.

The bound counterion free energy is approximated by
\begin{equation}
\beta f_{\rm bound}\simeq (Z-\tilde Z)\left[\ln\left(\frac{Z-\tilde Z}{v_s}\Lambda^3\right)-1\right]
+\frac{\tilde Z^2\lambda_B}{2a}~,
\label{fbound}
\end{equation}
the first term on the right side being the ideal-gas free energy of the bound counterions
in the association shell of volume $v_s=(4\pi/3)[(a+\delta)^3-a^3]$ and the second term
accounting for the self energy of a dressed macroion, assuming the bound counterions
to be localized near the macroion surface ($r=a$).

Given a bare valence, the effective valence equalizes the chemical potentials of counterions
in the free and bound phases.  This condition is equivalent to minimizing the volume energy
[Eq.~(\ref{epsilonv})] at fixed temperature and mean microion
densities~\cite{levin98,tamashiro98,levin01}:
\begin{equation}
\left(\frac{\partial\varepsilon_v}{\partial \tilde Z}\right)_{T,\bar n_{\pm}}=0
\label{Z}
\end{equation}
with $\tilde Z$ and $\delta$ related by Eq.~(\ref{delta2}).  The resultant effective valence and
corresponding shell thickness in turn determine the effective screening constant $\tilde\kappa$.

The effective pair potential between dressed macroions is given in terms of
the effective valence and screening constant:
\begin{equation}
\beta\tilde v_{\rm eff}(r)=\tilde Z^2\lambda_B\left(\frac{e^{\tilde\kappa a}}
{1+\tilde\kappa a}\right)^2\frac{e^{-\tilde\kappa r}}{r}~,\quad r>2(a+\delta)~,
\label{veffr}
\end{equation}
from which the macroion free energy can be approximated as [cf.~(\ref{Fm2})]
\begin{equation}
\hspace*{-2.5cm}
f_m(n_m,\tilde n_{\pm})=\min_{(d)}\left\{f_{\rm hs}(n_m,\tilde n_{\pm};d)
+2\pi n_m\int_d^{\infty}{\rm d}r\,r^2g_{\rm hs}(r,n_m;d)
\tilde v_{\rm eff}(r,n_m,\tilde n_{\pm})\right\}~.
\label{fm}
\end{equation}
The hard-sphere fluid functions, $f_{\rm hs}$ and $g_{\rm hs}$, are calculated
from the very accurate Carnahan-Starling and Verlet-Weis expressions~\cite{HM},
respectively.  In practice, the renormalized system parameters ($\tilde Z$, $\delta$,
$\tilde\kappa$) must be held constant in the variational minimization and in
all thermodynamic partial derivatives.

From the total semigrand potential per macroion, $\omega=\varepsilon_v+f_m$,
the thermodynamic pressure finally can be calculated via
\begin{equation}
p=n_m^2\left(\frac{\partial\omega}{\partial n_m}\right)_{T,n_0}
=p_{\rm free}+p_m~,
\label{ptot}
\end{equation}
where
\begin{equation}
\beta p_{\rm free}=\tilde n_++\tilde n_-
-\frac{\tilde Z(\tilde n_+-\tilde n_-)\tilde\kappa\lambda_B}{4[1+\tilde\kappa(a+\delta)]^2}
\label{pfree}
\end{equation}
is the (reduced) pressure due to the entropy and energy of the free microions
(bound counterions do not contribute directly, since $\tilde Z$ and $\delta$ are
implicitly fixed in the partial derivatives) and
\begin{equation}
\beta p_m=n_m+n_m^2\beta\left(\frac{\partial f_m}{\partial n_m}\right)_{T,n_0}
\label{pm}
\end{equation}
is the contribution from the entropy and effective macroion interactions.
The macroion pressure also can be obtained from computer simulation using the
virial theorem, properly generalized to account for the density dependence of
the effective pair potential~\cite{lu-denton07}.

Application of charge renormalization theory to bulk suspensions requires setting
the parameter $C$ in Eqs.~(\ref{delta1}) and (\ref{delta2}) to define the
threshold for renormalization at which bare and effective macroion valences begin
to differ.  Salt-free suspensions afford some freedom to set this threshold.
In previous applications~\cite{denton08,lu-denton10}, the choice $C=2$
yielded predictions in close agreement with primitive model simulation data for
thermodynamic and structural properties of deionized suspensions~\cite{note1}.
For salty suspensions, however, the physical requirement that the coion density
be non-negative [$n_-({\bf r})\ge~0$ in Eq.~(\ref{npm-lin})] constrains $C\leq~1$.
Moreover, the presumed exclusion of coions from the association shell
(Sec.~\ref{CRM}) further dictates that $C\geq~1$.  Thus, the renormalization threshold
is uniquely defined by setting $C=1$ in the present generalized theory.

\section{Results and Discussion}\label{results}

We consider the biologically relevant case of aqueous suspensions at room temperature
($\lambda_B=0.72$ nm) and restrict attention to monovalent microions, justifying
neglect of microion correlations in the mean-field Poisson-Boltzmann theory.
Three different implementations of PB theory are here compared:
(1) the full nonlinear PB theory in a spherical cell model (PB-cell);
(2) linearized PB theory in the cell model with effective (renormalized) valence
(LPB-cell); and (3) linearized PB theory in the effective-interaction model
(no cell boundaries) with effective valence (LPB-EI).
The nonlinear PB equation with spherical cell boundary conditions [Eq.~(\ref{PB2})]
was solved numerically by adapting the elegant Mathematica algorithm proposed by
Trizac {\it et al.}~\cite{trizac-langmuir03}.  For the linearized PB equation with
cell boundary conditions, the electrostatic potential was obtained analytically from
Eqs.~(\ref{LPB-cell})-(\ref{A3}) and the pressure numerically from
Eq.~(\ref{pressure-omega}).  For the effective-interaction model, the macroion
free energy was calculated numerically from Eqs.~(\ref{veffr}) and (\ref{fm})
and the pressure from Eqs.~(\ref{ptot})-(\ref{pm}).

Beginning with salt-free suspensions, Figs.~\ref{fig-pblincr} and
{\ref{fig-castaneda-priego} pit PB theory against primitive model
simulations~\cite{linse00,castaneda-priego06,linse00-note}.
Predictions are compared for the equation of state (pressure or osmotic coefficient)
at various electrostatic coupling strengths $\Gamma=\lambda_B/a$.
Figure~\ref{fig-pblincr} shows results for $Z=40$ and $0.0222<\Gamma<0.7115$,
and Fig.~{\ref{fig-castaneda-priego} for $Z=60$ and $\Gamma=0.3245$.
All three implementations of PB theory are seen to agree closely with simulation,
even well above the renormalization threshold ($Z\Gamma>5$),
the parameter regime of highly charged latex particles and ionic surfactant micelles.
The renormalized linearized PB models prove comparable in accuracy to
the renormalized jellium model~\cite{castaneda-priego06,schurtenberger08}.
Previous work~\cite{lu-denton10,lu-denton07} confirmed the high accuracy of the
variational approximation for the macroion free energy [Eq.~(\ref{fm})] by
comparing the LPB-EI model to Monte Carlo simulations of the one-component model
with the same effective interactions.

Figure~\ref{fig-deserno} compares theoretical predictions for the equation of
state of a highly deionized suspension at low, but non-zero, salt concentration.
For this strongly coupled system ($Z\Gamma\simeq~19$), the effective macroion
valence is significantly lower than the bare valence (inset).  The same three
implementations of PB theory give close mutual agreement for the osmotic pressure.
In contrast, linearized PB theory with bare macroion valence (no renormalization)
yields qualitatively different predictions, including {\it negative} osmotic
pressure at lower volume fractions.  This unphysical behaviour confirms previous
findings~\cite{denton08,lu-denton10} that charge renormalization can profoundly
affect thermodynamic properties.  When consistently combined with
renormalization theory, however, the cell and effective-interaction models are
comparably accurate in describing counterion-dominated suspensions.

Turning to suspensions in Donnan equilibrium with an electrolyte reservoir,
we explore the influence of reservoir salt concentration $n_0$ on osmotic pressure.
Figure~\ref{fig-pcr-salt} compares predictions for the equation of state of a
suspension of highly charged macroions with $a=50$ nm, $Z=500$, over a wide range
of concentrations.  With increasing ionic strength, small relative differences in
osmotic pressure for $n_0<0.1$~mM grow to significant deviations for
$n_0>0.4$~mM (Fig.~\ref{fig-pcr-salt}a).
At concentrations in the range 1~mM $<n_0<10$~mM (Figs.~\ref{fig-pcr-salt}b
and~\ref{fig-pcr-salt}c), a sizeable gap separates the predictions of nonlinear
PB theory in the cell model and linearized PB theory in the (charge-renormalized)
effective-interaction model.  This gap can be traced to the free energy and
pressure contributions from microion-induced interactions and correlations
between macroions, which are naturally included in the effective-interaction model
but omitted from the cell model~\cite{dobnikar06,trizac07}.  Nevertheless,
the salt concentrations predicted by the PB-cell and LPB-EI models are in
near-exact agreement (inset to Fig.~\ref{fig-pcr-salt}c).

At reservoir concentration $n_0=1$~mM, the cell-model implementation of
linearized PB theory predicts negative osmotic pressure and bulk modulus
over a range of volume fractions.  However, since the nonlinear PB-cell model
predicts strictly positive pressure and
bulk modulus~\cite{deserno02,trizac-jcp03}, as does the LPB-EI model, this
anomalous prediction can only be an artifact of the LPB-cell model at intermediate
ionic strengths.  At physiological and higher salt concentrations ($n_0>100$~mM),
microion screening so weakens effective macroion interactions that accuracy
of the cell model is restored for the colloidal parameters here studied.

Finally, we revisit the fundamental issue of thermodynamic phase stability.
Previous studies based on linearized PB theory~\cite{vRH97,graf98,vRDH99,vRE99,warren00} 
and nonlinear response theory~\cite{denton06}, without regard for charge renormalization, 
predicted a spinodal instability in deionized suspensions at low (but nonzero) salt concentrations.
Separation into macroion-rich (liquid) and macroion-poor (vapour) phases was
predicted in a parameter regime ($Z\lambda_B/a>7$) that lies slightly beyond
the threshold for charge renormalization.  Subsequent
analysis~\cite{vongruenberg01,klein01,deserno02,tamashiro03,trizac-jcp03}
attributed this unusual instability to an artifact of linearization approximations.
The theory of Zoetekouw and van Roij~\cite{zoetekouw_prl06,zoetekouw_pre06},
which incorporates an effective macroion valence, yields a revised prediction of
instability at much stronger couplings ($Z\lambda_B/a>24$).  This prediction,
however, being deduced from a mean-field theory, cannot be directly compared with
simulations of salt-free suspensions~\cite{linse-lobaskin99,linse00,brukhno09,
messina00,lobaskin-qamhieh03,rescic-linse01} that exhibit macroion aggregation
driven by counterion correlations.

In contrast, the present effective-interaction implementation of linearized PB theory,
surveyed over wide ranges of colloidal parameters ($a$, $Z$, $\eta$) and salt
concentrations, predicts no phase instability for any physically rational
renormalization threshold ($C\leq~1$).  Instability emerges only when the theory is
pushed beyond unphysical renormalization thresholds ($C>1$) that imply negative
coion densities.  The absence of thermodynamic instability predicted here is
consistent with the renormalized jellium model~\cite{levin03,trizac-levin04},
the cell-model implementation of nonlinear PB theory, which admits only a
positive bulk modulus~\cite{deserno02,trizac-jcp03}, and with primitive model
simulations for weakly correlated microions~\cite{brukhno-private}.
The apparent discrepancy between predictions of the present approach and that of
refs.~\cite{zoetekouw_prl06,zoetekouw_pre06} could be related to the different
reference potentials chosen for linearization of the PB equation.

\section{Conclusions}\label{conclusions}

Summarizing, a theory of charge renormalization that was previously developed for
salt-free colloidal suspensions is here generalized to suspensions dialyzed against
an electrolyte reservoir.  The theory incorporates into the one-component model an
effective macroion valence, defined via a simple thermal criterion.  Effective
electrostatic interactions --- one-body volume energy and effective pair potential
--- are derived from Poisson-Boltzmann theory, via a linear-screening approximation,
and depend on the renormalized parameters.  In contrast to the salt-free case,
the presence of salt ions imposes a unique threshold for charge renormalization.

When applied to salt-free, aqueous suspensions with monovalent counterions, the
renormalized effective-interaction implementation of linearized PB theory predicts
osmotic pressures in excellent agreement with the cell-model implementation of
nonlinear PB theory and with available data from Monte Carlo simulations of the
primitive model.  With increasing salt concentration, predictions of the
effective-interaction and cell-model implementations increasingly deviate from
one another.  The deviations are attributed to the relatively large pressure
contribution originating from microion-induced effective interactions and correlations
between macroions, which are neglected in the cell-model implementation.  At physiological
and higher salt concentrations, however, the macroions are so strongly screened
that the relative deviations are negligible.  We conclude that the cell-model
implementation of PB theory, although remarkably accurate at relatively low and high
salt concentrations, is less reliable at intermediate salt concentrations,
roughly in the range 0.5~mM$<c_s<$50~mM.  More extensive data from controlled
experiments and simulations of the primitive model, especially for bulk suspensions
at significant salt concentrations, are required to test the accuracy of competing theories.

The present approach reveals no indication of a liquid-vapour phase instability.
Nevertheless, counterion screening in deionized suspensions undoubtedly promotes
macroion cohesion, as manifested in the density dependence of the volume energy.
Therefore, effective electrostatic interactions, if not driving, may nevertheless
assist an instability that is favoured also by other cohesive mechanisms, such as
attractive van der Waals or depletion interactions.  The complex interplay between
such complementary mechanisms for phase separation will be explored in future work.

\ack
Helpful discussions with Sylvio May and correspondence with
Andrey Brukhno, Ramon Casta\~neda-Priego, Marcus Deserno, and Hans-Hennig von Gr\"unberg
are gratefully acknowledged.
This work was partially supported by the National Science Foundation (DMR-0204020).
Acknowledgment is made to the Donors of the American Chemical Society Petroleum
Research Fund (PRF 44365-AC7) for partial support of this research.

%\appendix*

%\clearpage
%\begin{references}

\newpage

\begin{figure}[h!]
\vspace*{2cm}
\includegraphics[width=0.85\columnwidth]{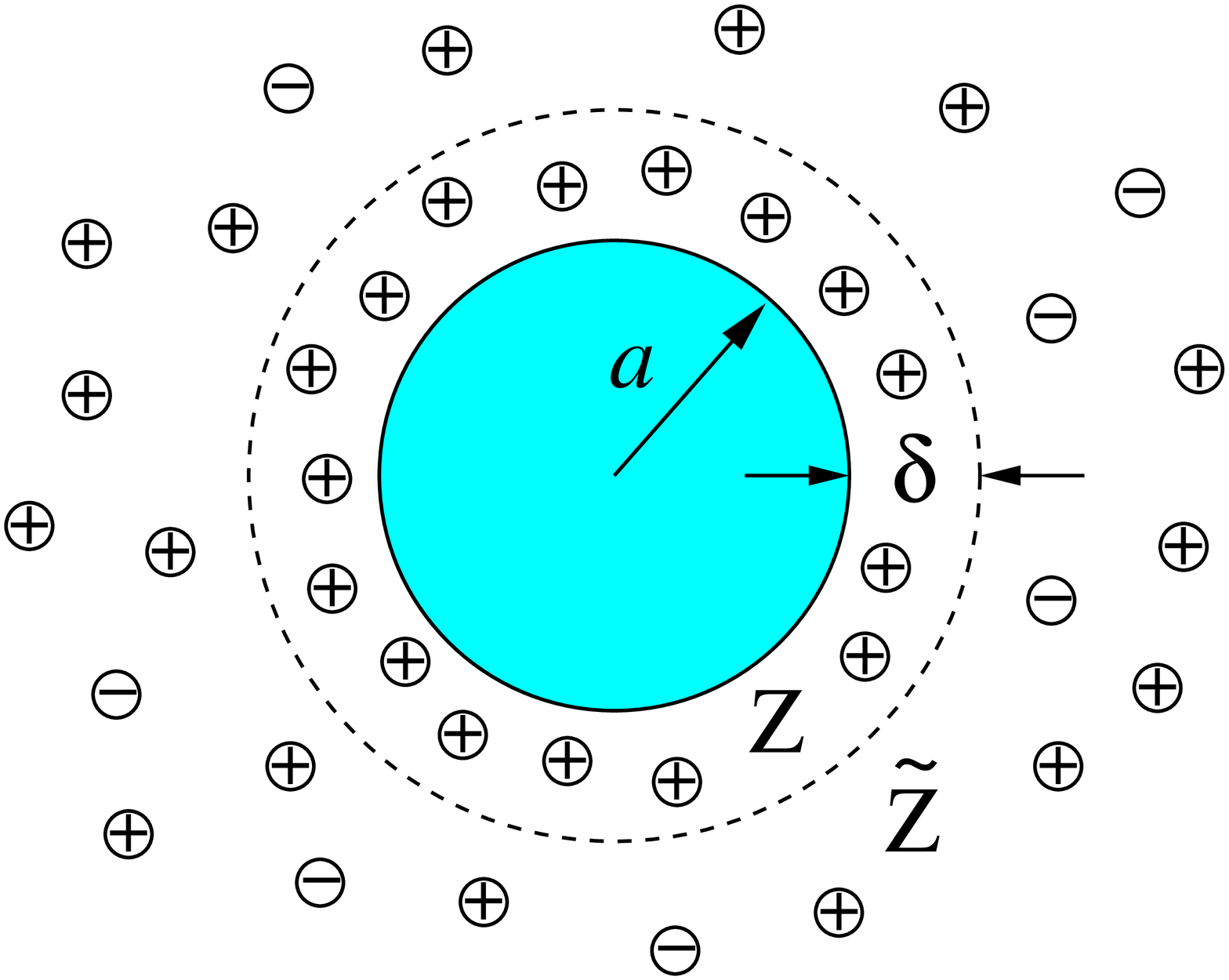}
\caption{\label{fig-model}
Model of charged colloidal suspension: spherical macroions of radius $a$
and point microions dispersed in a dielectric continuum.  Strongly associated
counterions in a spherical shell of thickness $\delta$ renormalize the
bare macroion valence $Z$ to an effective (reduced) valence $\tilde Z$.
}
\end{figure}

\newpage
\begin{figure}
\vspace*{2cm}
\includegraphics[width=0.85\columnwidth]{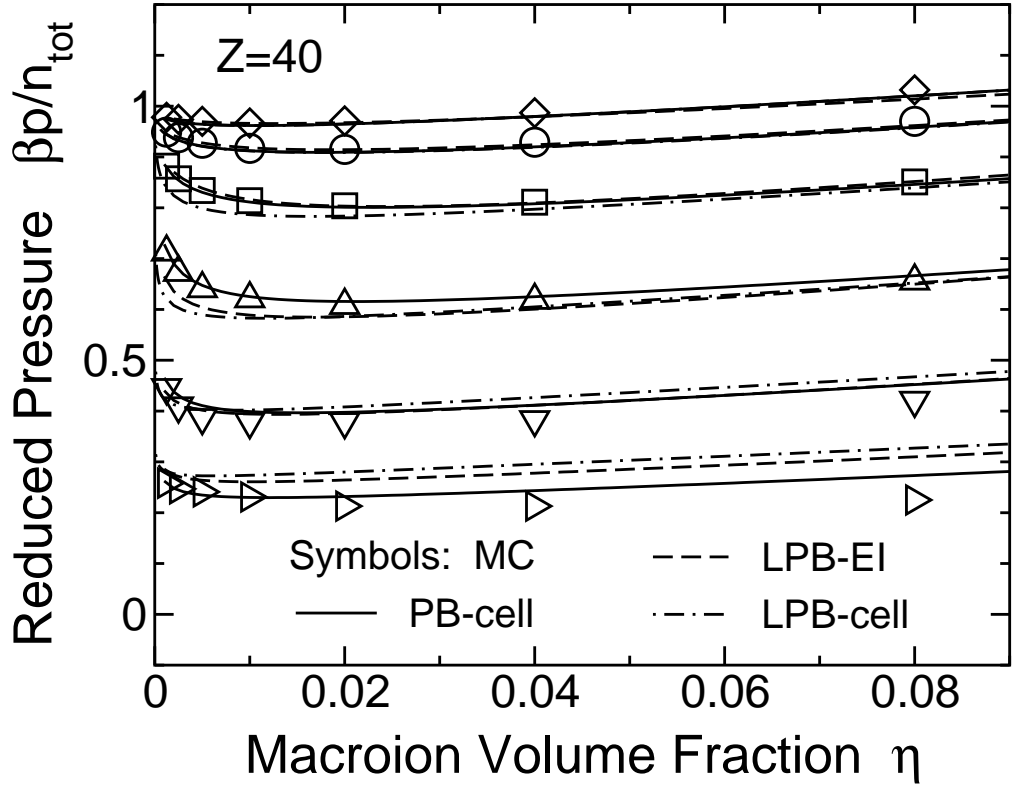}
\caption{\label{fig-pblincr}
Reduced pressure $\beta p/n_{\rm tot}$ vs.~macroion volume fraction $\eta$,
where $n_{\rm tot}=(Z+1)n_m$ (total ion density), of salt-free suspensions
with bare macroion valence $Z=40$ and electrostatic coupling strengths
(top to bottom) $\Gamma=\lambda_B/a=0.0222$, 0.0445, 0.0889, 0.1779, 0.3558, 0.7115.
Symbols: Monte Carlo simulations of the primitive model~\cite{linse00}
(symbol sizes exceed error bars).  Curves: Poisson-Boltzmann cell model (PB-cell, solid),
linearized Poisson-Boltzmann cell model (LPB-cell, dot-dashed), and
linearized effective-interaction theory (LPB-EI, dashed).
For $\Gamma>0.1$, the effective macroion charge is renormalized ($\tilde Z<Z$).
}
\end{figure}

\newpage
\begin{figure}
\vspace*{2cm}
\includegraphics[width=0.85\columnwidth]{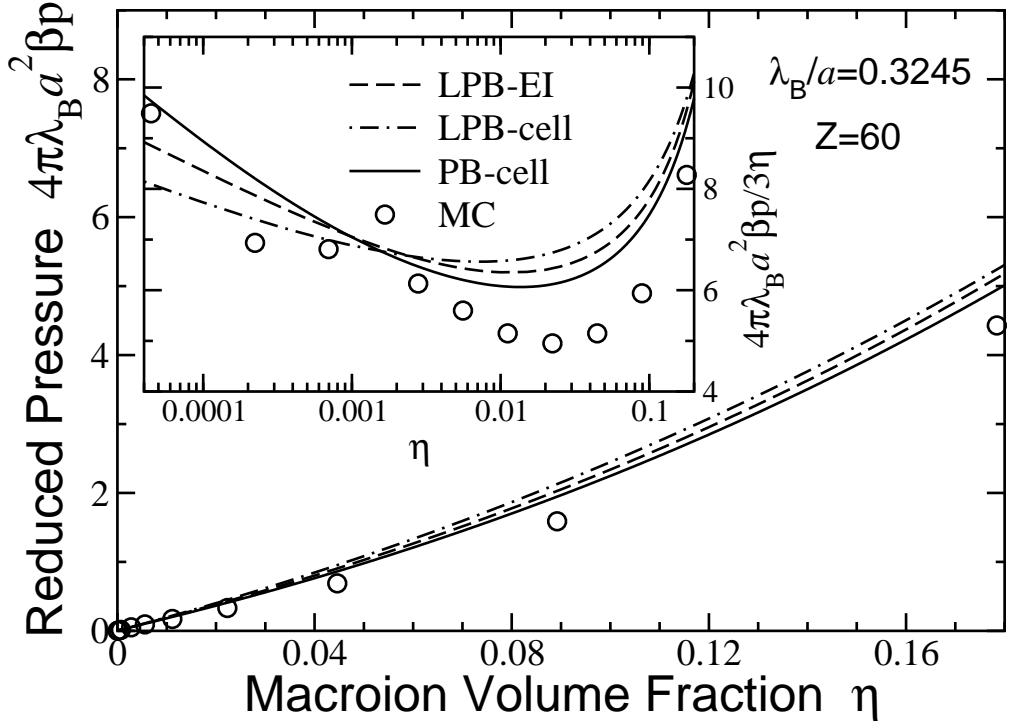}
\vspace*{0.5cm}
\caption{\label{fig-castaneda-priego}
Reduced pressure vs.~volume fraction $\eta$ of a salt-free suspension with bare macroion
valence $Z=60$ and electrostatic coupling strength $\Gamma=\lambda_B/a=0.3245$.
Symbols: Monte Carlo simulations of the primitive model~\cite{castaneda-priego06}.
Curves: Poisson-Boltzmann cell model (PB-cell, solid),
linearized Poisson-Boltzmann cell model (LPB-cell, dot-dashed), and
linearized effective-interaction theory (LPB-EI, dashed).
Inset: osmotic coefficients over several decades of volume fraction.
}
\end{figure}

\newpage
\begin{figure}
\vspace*{2cm}
\includegraphics[width=0.85\columnwidth]{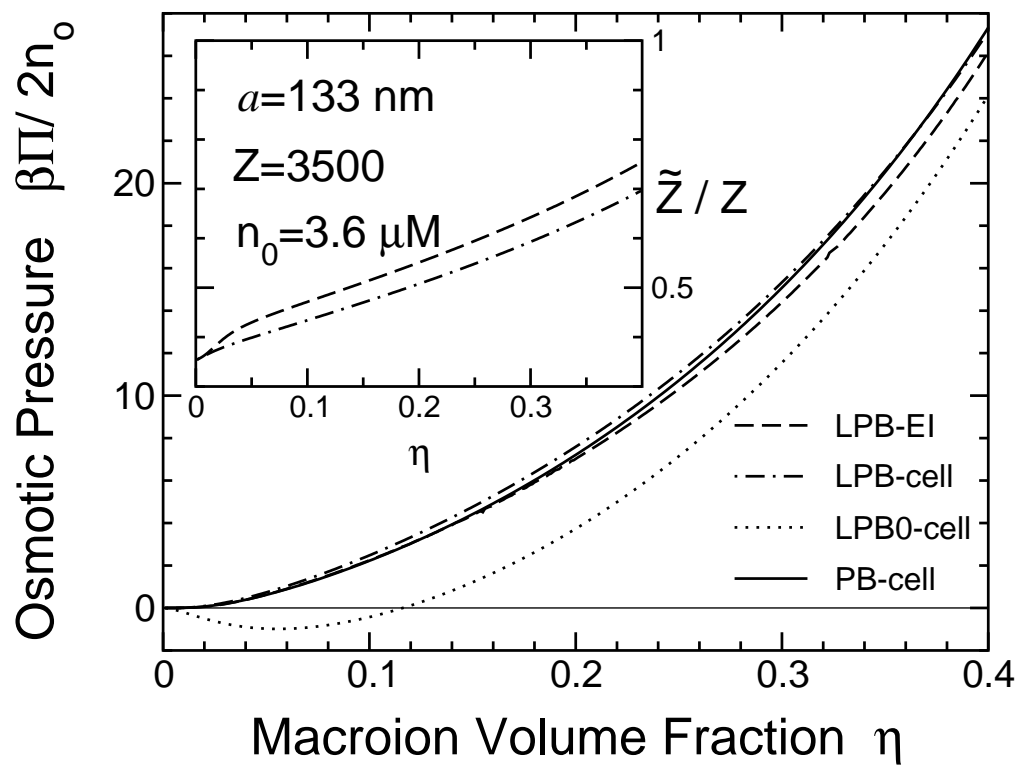}
\caption{\label{fig-deserno}
Reduced osmotic pressure vs. volume fraction of a suspension with macroion radius
$a=133$ nm and bare valence $Z=3500$ in Donnan equilibrium with a salt reservoir of
concentration $n_0=3.6~\mu$M.  Predictions of alternative theoretical implementations
of Poisson-Boltzmann theory are compared:
linearized PB theory in the effective-interaction (LPB-EI, dashed) and
cell-model (LPB-cell, dot-dashed) implementations, both incorporating charge renormalization;
linearized cell model without charge renormalization (LPB0-cell, dotted);
nonlinear PB theory in the cell model (PB-cell, solid).
}
\end{figure}

\newpage
\begin{figure}
\vspace*{2cm}
\includegraphics[width=0.85\columnwidth]{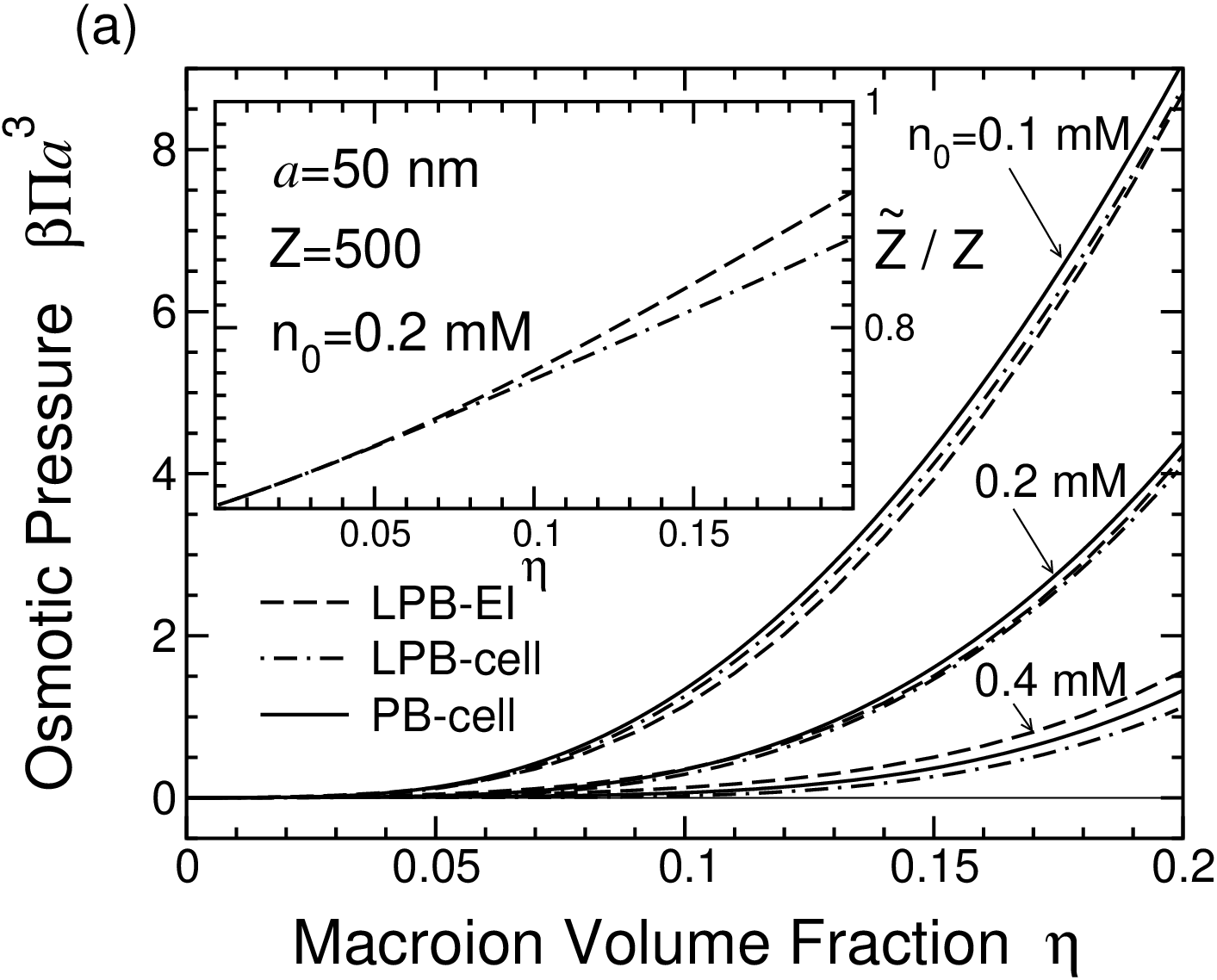} \\[6ex]
\includegraphics[width=0.85\columnwidth]{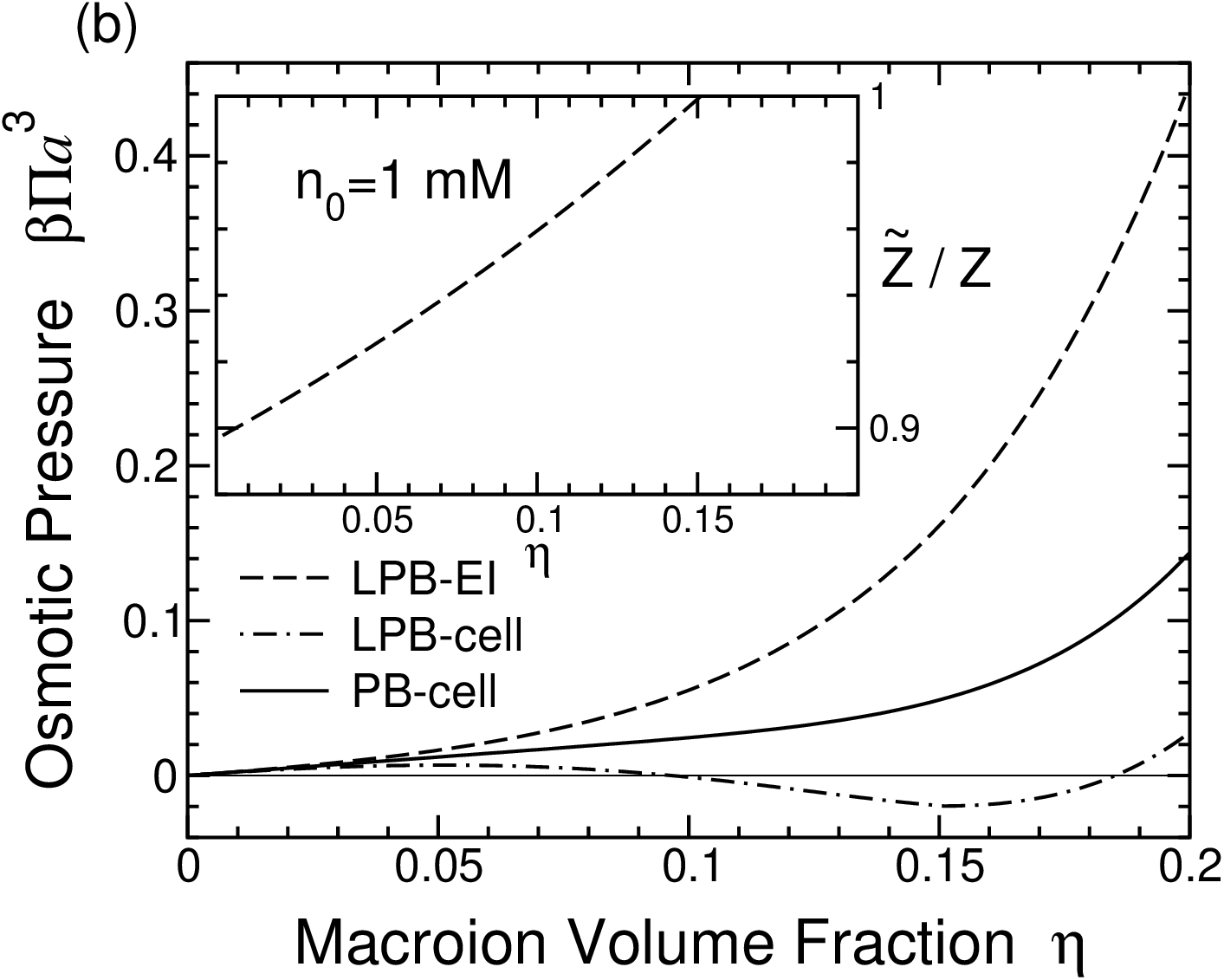}
\end{figure}

\newpage
\begin{figure}
\vspace*{2cm}
\includegraphics[width=0.85\columnwidth]{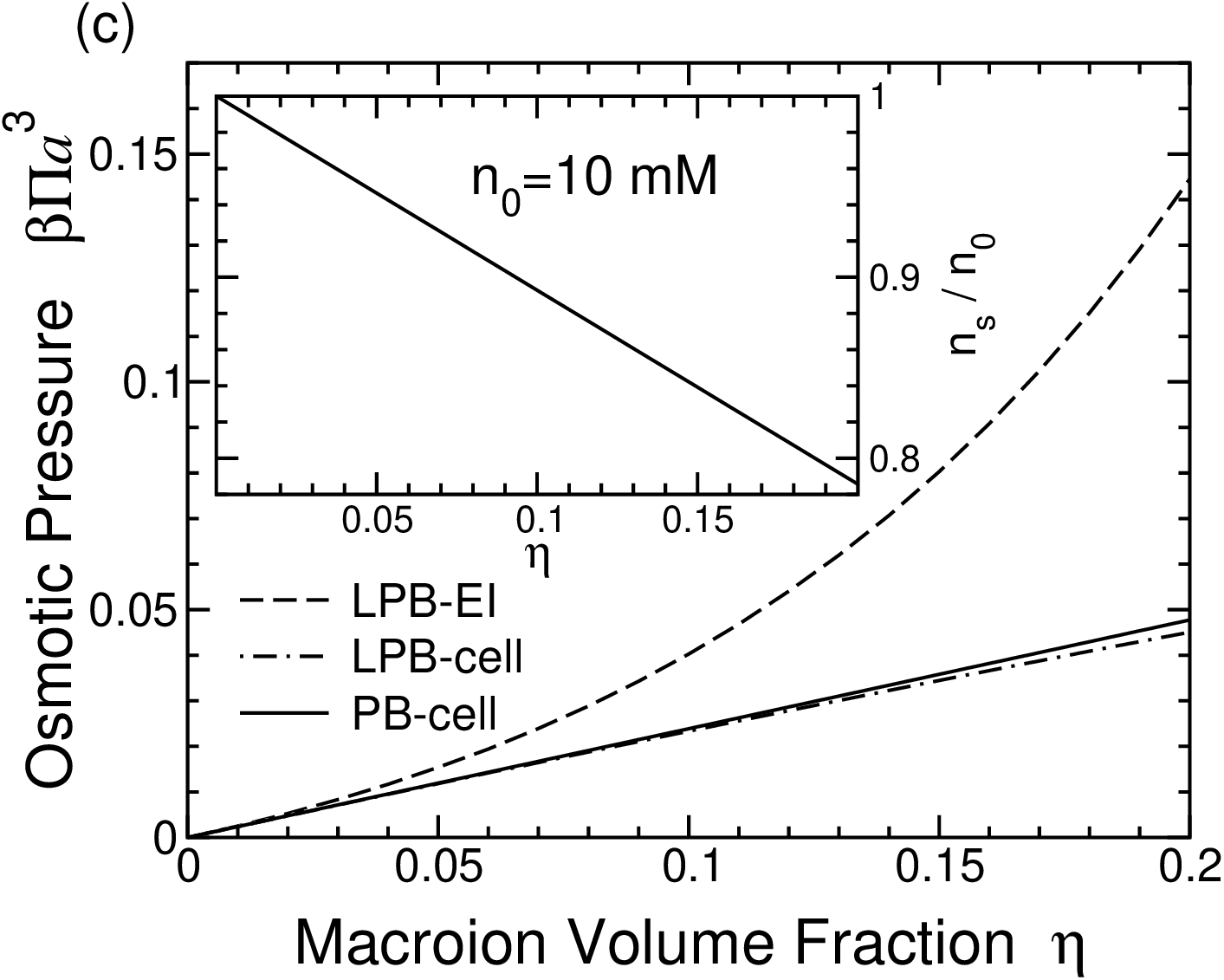}
\caption{\label{fig-pcr-salt}
Reduced osmotic pressure vs. volume fraction of a suspension with macroion radius
$a=50$ nm and bare valence $Z=500$ in Donnan equilibrium with a salt reservoir of
various concentrations: $n_0=0.1, 0.2, 0.4$~mM (a), 1~mM (b), and 10~mM (c).
Predictions of alternative implementations of Poisson-Boltzmann theory are compared:
linearized PB theory in the effective-interaction (LPB-EI, dashed) and
cell-model (LPB-cell, dot-dashed) implementations;
nonlinear PB theory in the cell model (PB-cell, solid).
Inset to panels (a) and (b): ratio of effective to bare valence
[predictions of the two LPB models are indistinguishable in panel (b)].
Inset to panel (c): ratio of salt concentrations in suspension and reservoir
(predictions of the three models are indistinguishable on this scale).
}
\end{figure}


\begin{thebibliography}{99}
\bibitem{note1} In salt-free suspensions, the mean reduced potential
[Eq.~(\ref{psibar-open-cr})] is $\langle\tilde\psi\rangle=-1$.
The choice $C=2$ in Eqs.~(\ref{delta1}) and (\ref{delta2}), as assumed in
refs.~\cite{denton08,lu-denton10}, then implies a renormalization threshold
potential $\tilde\psi(a+\delta)=-3$ relative to the reservoir,
or $|\tilde\psi(a+\delta)-\langle\tilde\psi\rangle|=2$ relative to the mean.
The present theory dictates $C=1$, which proves to maintain comparable accuracy
in the salt-free case.
\bibitem{linse00-note}
Note that reduced pressures in Table III of ref.~\cite{linse00} should be
$\beta p/n_{\rm tot}=0.361$ for $Z=10$, $\Gamma=1.423$, $\eta=0.01$;
0.970 for $Z=40$, $\Gamma=0.0222$, $\eta=0.005$; and
0.971 for $Z=40$, $\Gamma=0.0222$, $\eta=0.02$.
P.~Linse, private communication (2005).
\end{thebibliography}
\end{document}